\documentclass[a4paper,11pt,twoside]{scrartcl}
\usepackage{ILD}

\usepackage{hyperref}
\usepackage{url}
\hypersetup{colorlinks=true,citecolor=blue,urlcolor=blue,linkcolor=blue}
\graphicspath{{./logos/} {./plots/} }
\usepackage{xspace}

\newcommand{\whizard}{\textsc{Whizard}\xspace}
\newcommand{\pythia}{\textsc{Pythia}\xspace}
\newcommand{\delphes}{\textsc{Delphes}\xspace}
\newcommand{\epem}{\ensuremath{\textrm{e}^+\textrm{e}^-}\xspace}

\newlength{\figwidth}
\newlength{\twofigwidth}
\title{Prospects for light exotic scalar measurements\\ at the e$^+$e$^-$ Higgs factory}

\ildproc{PHYS}{2024}{004}

\date{September 29, 2024}

\addauthor{Bart{\l}omiej Brudnowski}{} 
\addauthor{Kamil Zembaczy{\'n}ski}{} 
\addauthor{Aleksander Filip \.Zarnecki}{} 

\addinstitute{}{Faculty of Physics, University of Warsaw, Poland}

\abstract{
The physics program of the Higgs factory will focus on measurements of
the 125 GeV Higgs boson, with the Higgs-strahlung process being the
dominant production channel at 250 GeV. However, production of extra
light scalars is still not excluded by the existing experimental data,
provided their coupling to the gauge bosons is sufficiently
suppressed. Fermion couplings of such a scalar could also be very
different from the SM predictions leading to non-standard decay
paterns. Presented in this contribution are results from the ongoing
studies on prospects of direct light scalar observation at future
Higgs factory experiments in different decay channels. 

\vspace*{1cm}

\begin{center}
  Presented at the International Workshop on Future Linear
  Colliders (LCWS 2024),\\ 8-11 July 2024, 
 contribution submitted to EPJ Web of Conferences.\\[2mm]
This work was carried out in the framework of the ILD concept group\\
  as a contribution to the ECFA e$^+$e$^-$ Higgs/EW/top factory study.
\end{center}
}

\begin{document}

\titlepage

\section{Motivation}

\begin{figure}[tb]
\centering
\includegraphics[width=\twofigwidth]{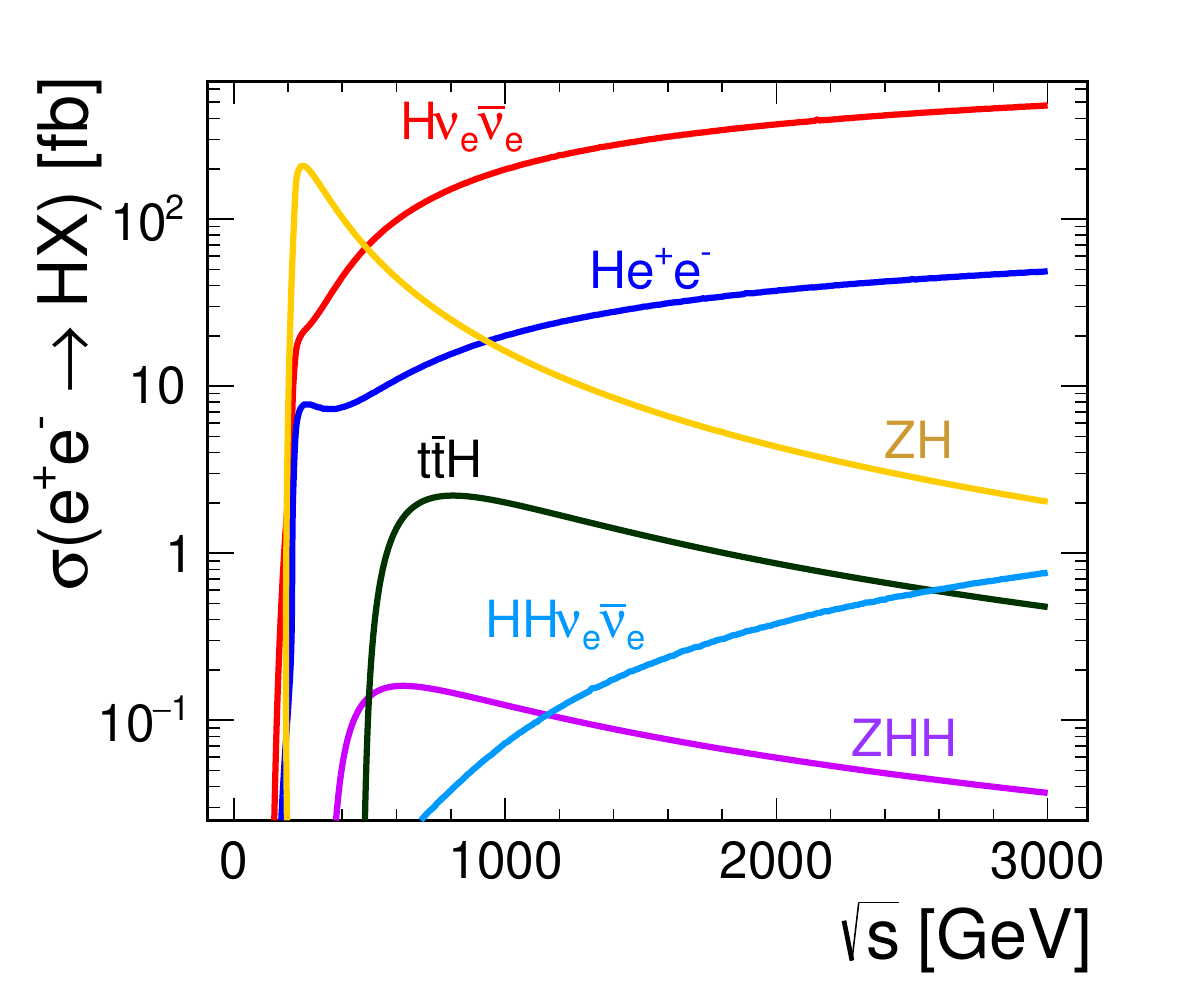}\hspace*{1cm}
\includegraphics[width=0.6\twofigwidth,trim=0 -6cm 0 0]{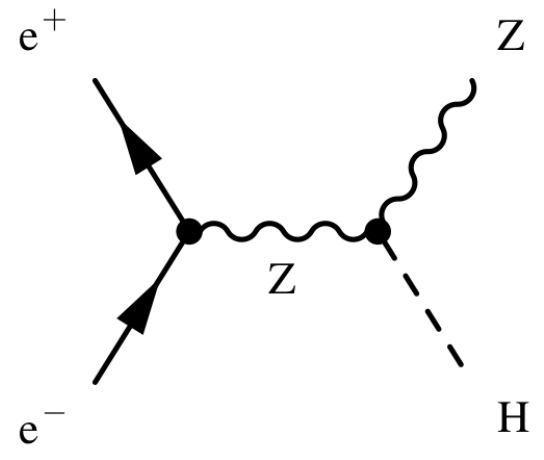}
\caption{Left: cross section as a function of centre-of-mass energy for the
  main Higgs production processes at an \epem collider. Right: the
  leading-order Feynman diagram for the Higgsstrahlung process.
  Figures taken from \cite{Abramowicz:2016zbo}.}
\label{higgs_prod}   
\end{figure}

In recent years, a general consensus was reached in the particle
physics community about the need for the next-generation large
infrastructure to be an electron-positron Higgs factory. It was indicated as the
highest-priority next collider in the 2020 Update of the  European
Strategy for Particle Physics~\cite{EuropeanStrategyGroup:2020pow}. 
While full exploitation of the Higgs boson physics requires running at
collision energies up to the TeV range, see
figure~\ref{higgs_prod}~(left), most of the precision measurements
will be carried out at the collision energy of 240--250\,GeV,
maximizing the cross section for the Higgs boson production in the so
called Higgsstrahlung process, see figure~\ref{higgs_prod}~(right).
However, existence of additional light scalar particles, with
masses of the order of or  below the mass of the 125\,GeV state
observed at the LHC, is  by far not excluded experimentally and also  
well motivated theoretically~\cite{Heinemeyer:2021msz,Biekotter:2022jyr,Robens:2022zgk}.
Higgs factories should be sensitive to exotic scalar production even
for very light scalars and small couplings, thanks to clean
environment, precision and hermeticity of the detectors. 
Still, prospects for light scalar measurements at the \epem Higgs
factory were hardly studied in the past.
That is why this subject was included as one of the so called focus
topics of the   ECFA e$^+$e$^-$ Higgs/EW/top factory study~\cite{deBlas:2024bmz}.
Two theoretical and phenomenological targets were defined:
associated production of the new scalar with the Z boson,
$\epem\to\;$Z\;S (scalar-strahlung process) and
light scalar pair-production in 125\,GeV Higgs boson decays, H\;$\to$\;S\;S.

\section{Decay mode independent search}

\begin{figure}[tb]
\centering
\includegraphics[width=1.03\twofigwidth]{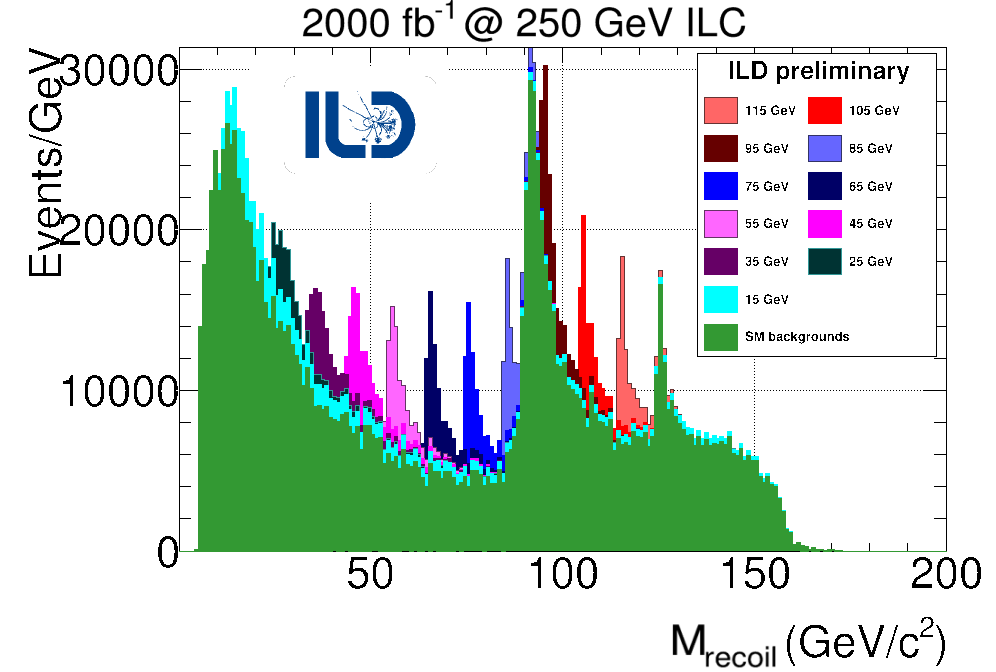}
\includegraphics[width=0.97\twofigwidth]{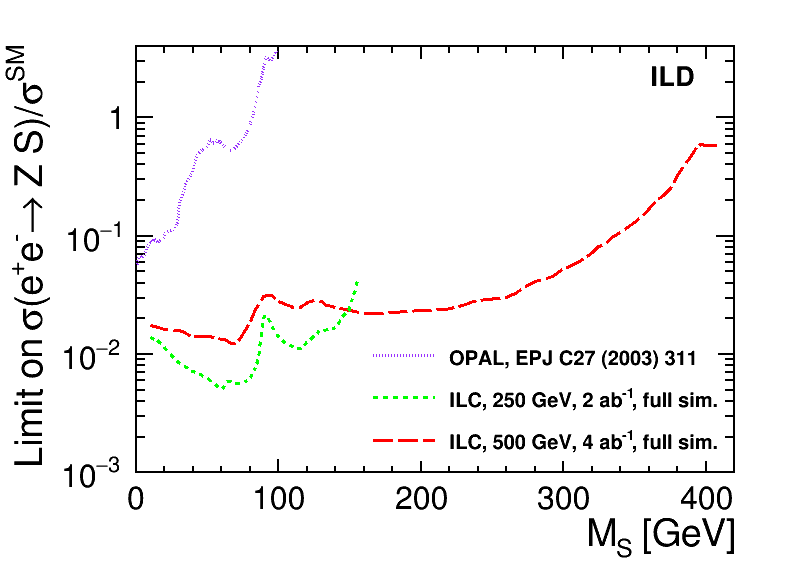}
\caption{Left: the recoil mass distributions after the selection cuts for signal of light scalar
  production and SM backgrounds processes for ILC running at 250\,GeV.
  Right: expected 95\% C.L. limits on the scalar production cross
  section, relative to the SM scalar production cross section at given
  mass, for ILC running at 250\,GeV and 500\,GeV~\cite{Wang:2018awp,Wang:2020lkq}.
}
\label{dec_ind}   
\end{figure}

As for the SM Higgs boson, the production of new scalars in the
scalar-strahlung process can be tagged, independent of their decay,
based on the recoil mass technique \cite{Yan:2016xyx}.  
For best recoil mass reconstruction Z decays to muon pair can be used,
which were exploited in the full simulation studies performed
within ILD~\cite{Wang:2018awp,Wang:2020lkq}.
Shown in figure~\ref{dec_ind}~(left) is the recoil mass distribution
expected for SM background processed at 250\,GeV ILC together with
expectations for different signal hypothesis.
Expected limits on the scalar production cross section,
relative to the SM scalar production cross section at given mass, are
compared with LEP limits based on the similar approach in figure~\ref{dec_ind}~(right).
Expected sensitivity of the experiment at future \epem Higgs factory
is an order of magnitude better than the existing LEP limit and the
mass range probed can be significantly extended as well.

\section{Search for S$\; \to \tau^+ \tau^-$}

\begin{figure}[tb]
  \centering
  \includegraphics[width=1.2\twofigwidth]{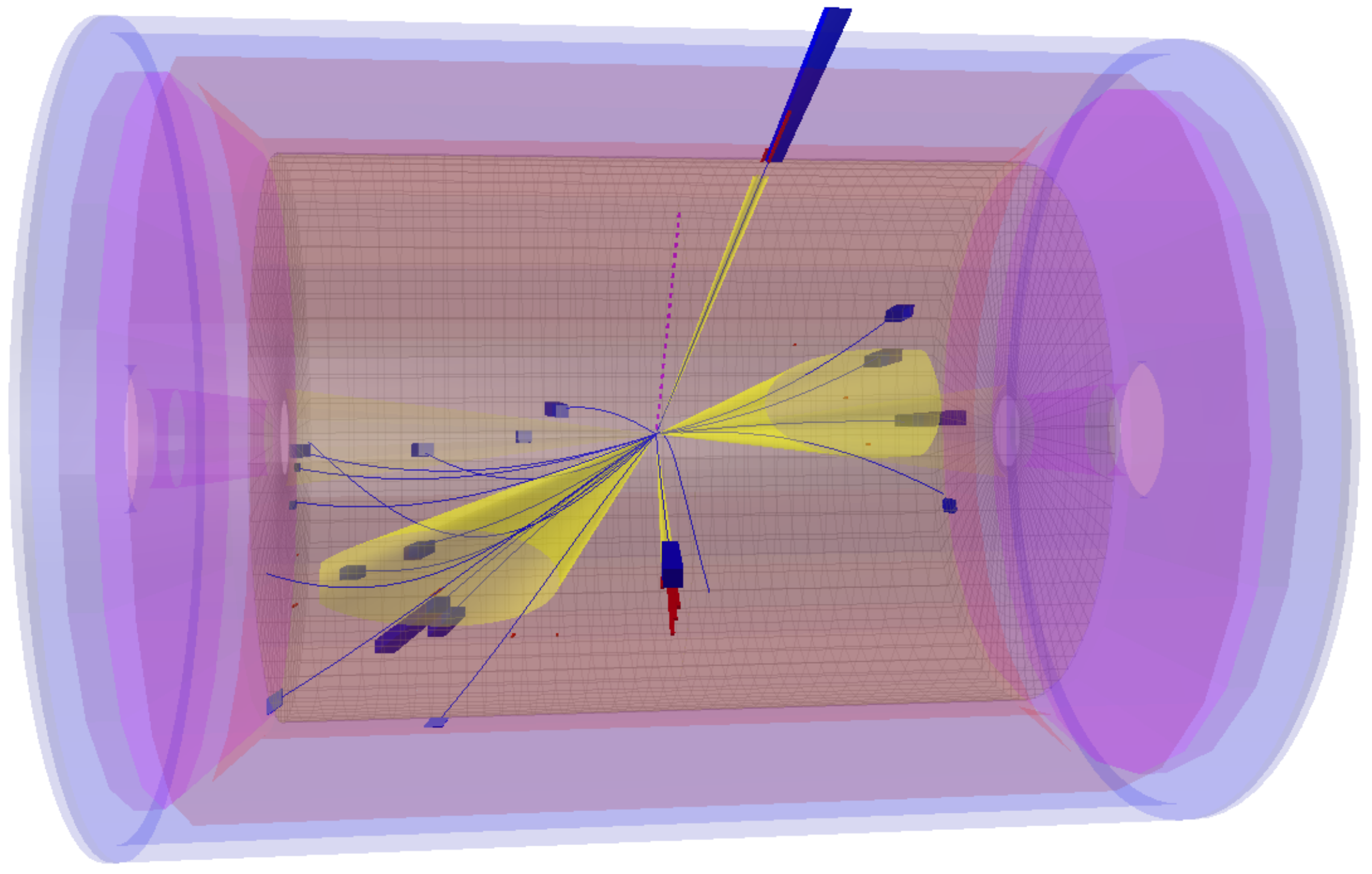}
  \includegraphics[width=0.8\twofigwidth]{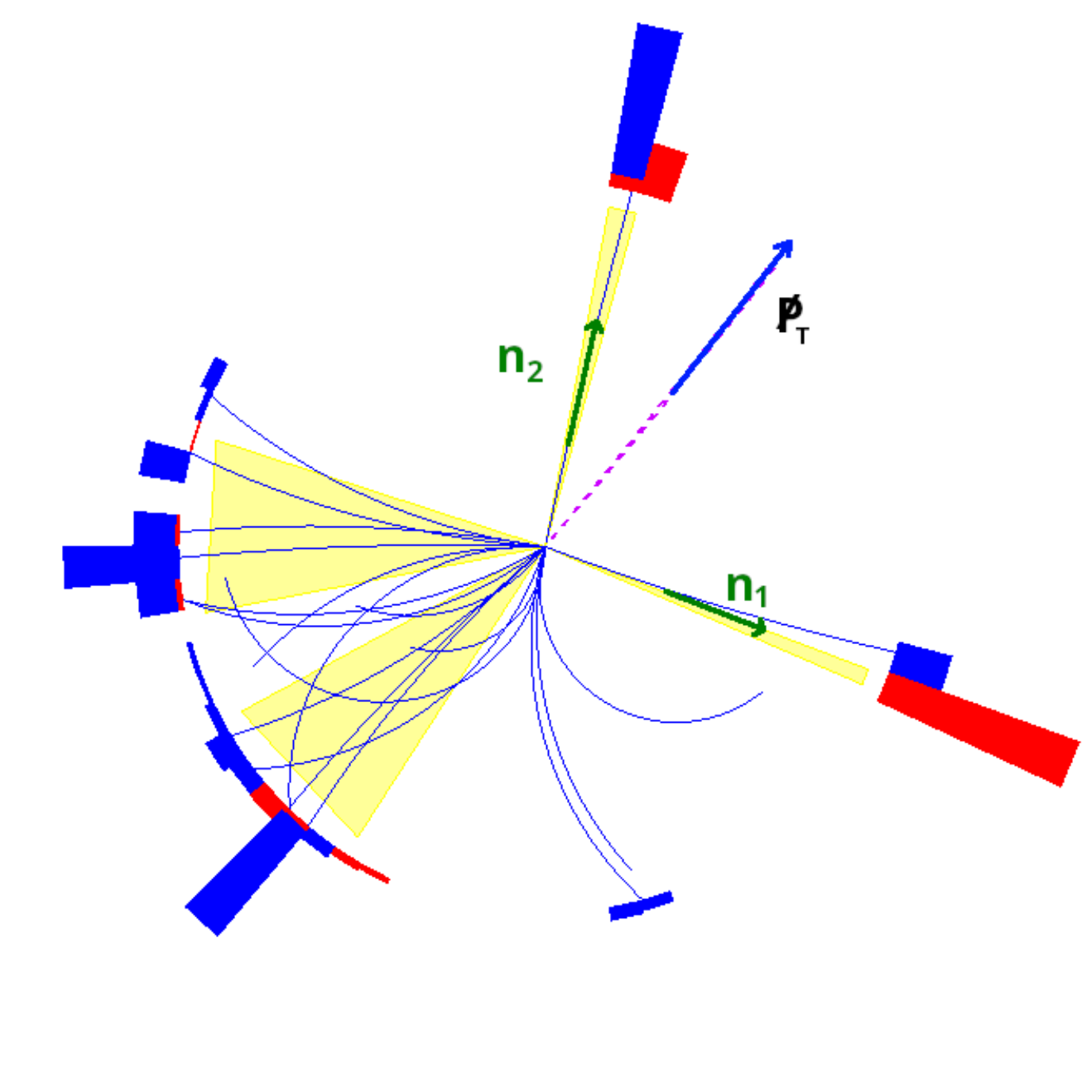}
  \caption{Left: example signal event, with hadronic decays of the two tau
  leptons produced in the light scalar decay. Right: same event in the
  transverse plane,
  missing transverse momentum
  $\vec{{p~}\!\!\!\!\!\!\!/}\!_{_{T}}$
  and two unit vectors along tau jet
  directions ($\vec{n}_1$ and $\vec{n}_2$) are indicated. }
\label{event}
\end{figure}

While limits resulting from the decay independent search based on the
recoil mass distribution only are the most general ones, significant
improvement of sensitivity is expected when particular decay channels
of the new scalars are addressed, at the price of some model dependence. 
As some of the discrepancies from SM predictions observed in LEP and LHC
data suggested possible existence of the new scalar with mass of about
95\,GeV and enhanced branching ratio to the $\tau^+ \tau^-$ final
state~\cite{Biekotter:2022jyr}, we decided to consider this decay channel as
the possible discovery scenario. 
Event samples used for the presented study were generated using
\whizard \cite{Moretti:2001zz,Kilian:2007gr} version 3.1.2.
Both background (including 125\,GeV Higgs boson production assuming
nominal couplings) and light scalar signal production were modeled
using  the built-in \texttt{SM\_CKM} model. 
%
As signal samples we generated Higgs boson production with subsequent
decays to tau lepton pair, for sets of modified Higgs masses.
Total lumionsity of 2\,ab$^{-1}$ was assumed for ILC running at
250\,GeV, as expected in the H-20 running scenario \cite{Bambade:2019fyw},
with $\pm80\%$ and $\pm30\%$ polarisation for electron and positron
beams, respectively.
The ILC beam energy profile, as simulated with GuineaPig \cite{Schulte:1999tx}, was
taken into account based on \textsc{Circe2} parametrisation and
hadronisation was simulated with the~\pythia~6~\cite{Sjostrand:2006za}. 
The fast detector simulation framework
\delphes~\cite{deFavereau:2013fsa} was used to simulate
detector response, with built-in cards for parametrisation of the ILC
detector, \texttt{delphes\_card\_ILCgen.tcl} \cite{bib:ILCgen}.
Example of signal event with hadronic final state, as simulated by \delphes, is shown in
figure~\ref{event}~(left).

Depending on the decays of the two tau leptons,
three decay channels can be considered for the signal events:
hadronic (with both taus decaying hadronically), semi-leptonic (with one
leptonic tau decay) and leptonic (with leptonic decays of both taus).
As a tight selection, we require each tau lepton to be identified
either as an isolated lepton (and missing $p_T$) or hadronic jet with
$\tau$-tag.
In addition to two tau candidates from the decay of the scalar, we also require
reconstruction of two (untagged) hadronic jets from the hadronic decay
of the Z boson.
However, as the efficiency of tau jet tagging implemented in \delphes
is relatively poor (at most 70\%), we also consider loose event
selection, when we require only one identified tau candidate (isolated
lepton or $\tau$-tagged jet) and three untagged hadronic jets, and
take the jet with smaller invariant mass as the second tau candidate.
At the pre-selection stage we thus select five event categories, as
summarised in table~\ref{evt_cat}.

One of the challenges in the search for scalar decays into tau leptons
is to properly reconstruct the invariant mass of the produced scalar, which can
be significantly underestimated due to the escaping neutrinos.
To correct for the neutrino energy, we use the so called collinear
approximation \cite{Kawada:2015wea}. For high energy tau leptons,
decay products are highly boosted in the initial lepton direction and one
can therefore assume that the initial tau lepton, escaping neutrino
and the observed tau candidate are collinear.
Neutrino energies can be found from transverse momentum balance:
\begin{equation}
 \vec{{p~}\!\!\!\!\!\!\!/}\!_{_{T}} \;= \; E_{\nu_1} \cdot \vec{n_1} + E_{\nu_2} \cdot \vec{n_2}
\end{equation}
where $\vec{n_1}$ and $\vec{n_2}$ are directions of the two tau
candidates in the transverse plane (see the right plot in
figure~\ref{event}).
While more advanced reconstruction methods exist, based on the
reconstruction of secondary decay vertex position, the advantage of
this method is that it can be applied to all events and the solution is unique.
\def\arraystretch{1.2}
\begin{table}[tb]
\centering
\caption{Event categories considered in the search for light scalar
  production with scalar decay to tau lepton pair.}
\label{evt_cat}       
        \begin{tabular}{|l|l|l|l|}
        \hline
         Event    &  Isolated  &  \multicolumn{2}{p{8cm}|}{Selection requirements}     \\ \cline{3-4}
         category &  leptons  &  tight selection &  loose selection \\ \hline
        hadronic &
        zero  & 4 jets, 2 with $\tau$-tag & 4 jets, 1 with $\tau$-tag  \\ 
        semi-leptonic & 
        one  & 3 jets, 1 with $\tau$-tag & 3 jets with no $\tau$-tag \\ 
        leptonic & 
        two  & two jets without $\tau$-tag &  \\ \hline
        \end{tabular}
        \end{table}

\begin{figure}[tb]
\centering
\includegraphics[width=\twofigwidth,clip]{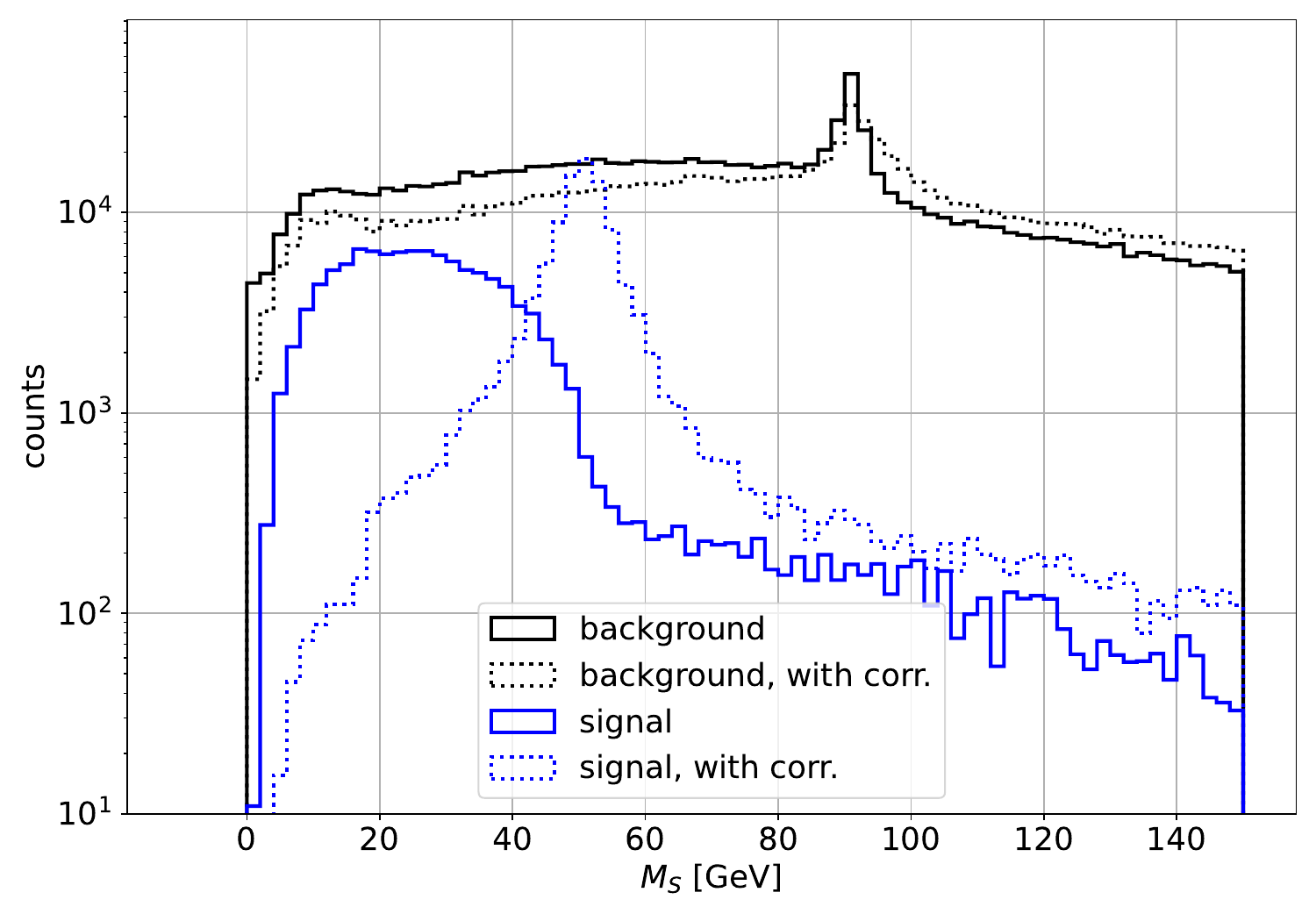}
\includegraphics[width=\twofigwidth,clip]{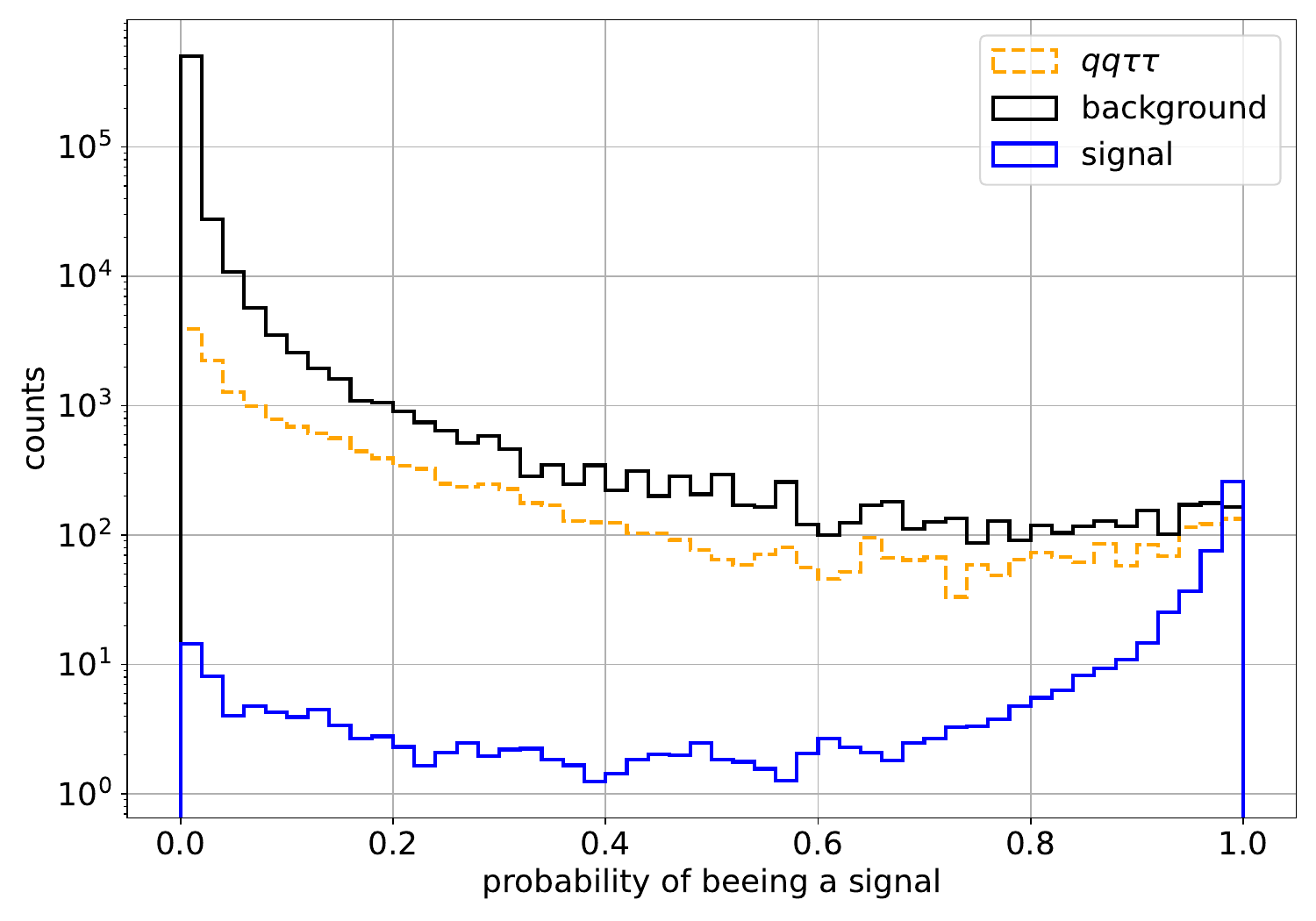}
\caption{Left: reconstructed invariant mass of the two tau candidates
  after tight selection for SM background (black) and signal of
  50\,GeV scalar production (blue) before (solid) and after (dashed) 
  collinear correction.
  Right: example of the BDT response distribution for 50\,GeV scalar signal (blue) and
  SM background (black) events, for tight semi-leptonic event
  selection and ILC running with $\textrm{e}^-_\textrm{L}\textrm{e}^+_\textrm{R}$
  polarisation combination.
  The signal cross section is normalized to 1\% of the SM Higgs boson
  production cross section at given scalar mass.
  Orange dashed lines indicated the contribution of the leading
  background process, $\epem \to \textrm{q} \textrm{q} \tau \tau $.
}
\label{tau_hist}   
\end{figure}

Shown in figure~\ref{tau_hist}~(left) are the mass distributions of
the tau candidate pairs in signal and background events, before
(solid) and after (dashed) collinear correction.
After the correction, the scalar mass can be reconstructed with about
5\,GeV precision, also for semi-leptonic and leptonic events.

For best event classification, we consider each event category (see
table~\ref{evt_cat}) and each beam polarisation combination separately,
resulting in 20 independent BDTs trained for event classification, for
each scalar mass considered.
Example of BDT response distribution is presented in
figure~\ref{tau_hist}.

\begin{figure}[tb]
\centering
\includegraphics[width=0.95\twofigwidth,clip]{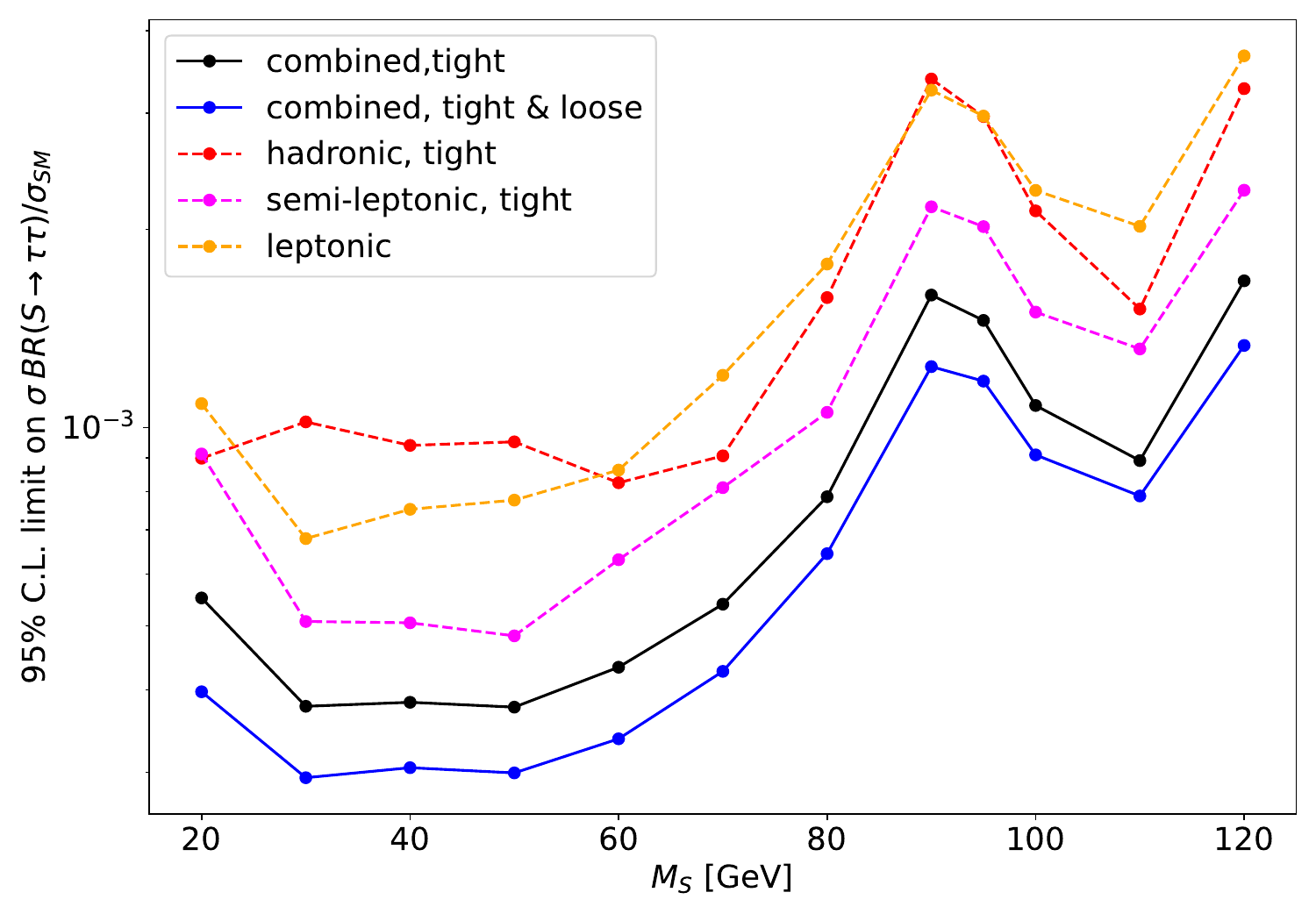}
\includegraphics[width=1.05\twofigwidth,trim=0 1cm 0 0]{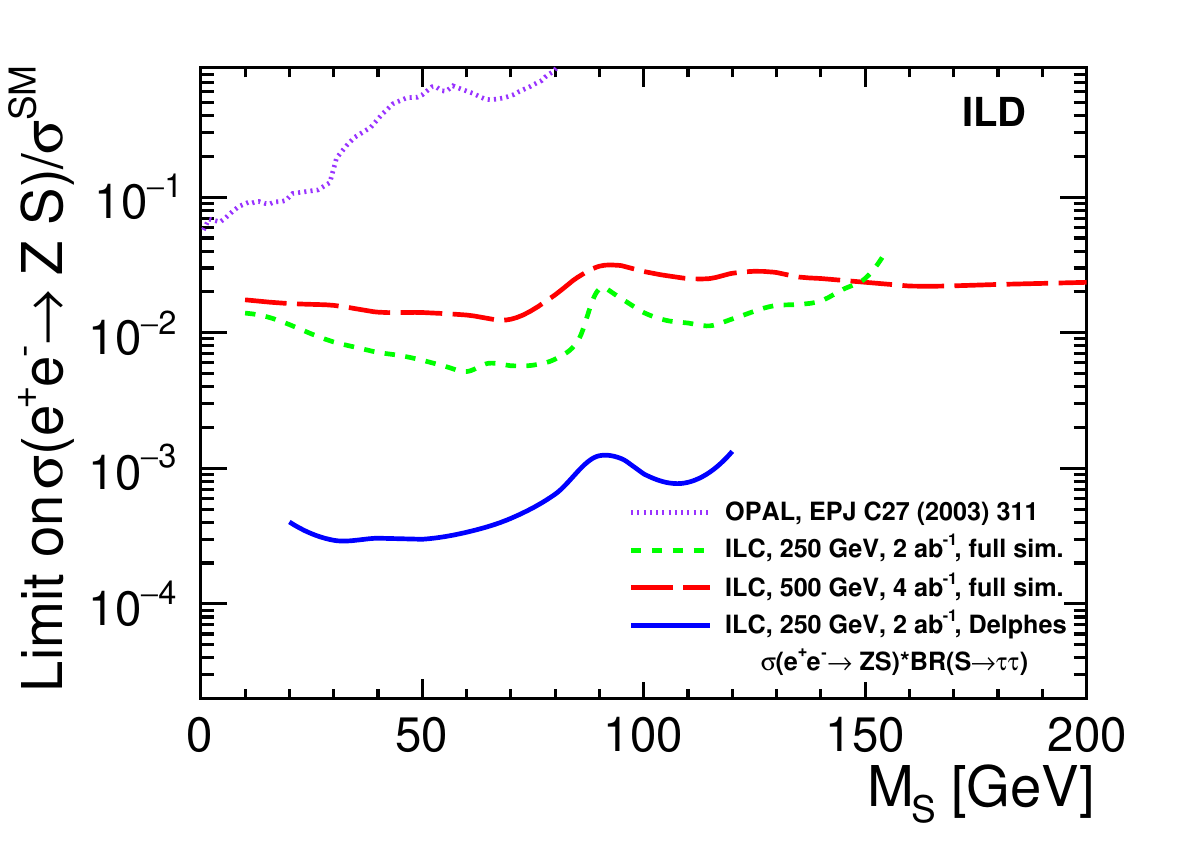}
\caption{Expected 95\% C.L. cross section limits on the light scalar
  production cross section times di-tau branching ratio for ILC
  running at 250 GeV.
  Left: comparison of combined limits with limits obtained for
  different event categories.
  Right: combined limits compared with limits resulting from decay independent
    study.}
\label{comb_ilc}   
\end{figure}

Expected exclusion limits, assuming no deviation from SM predictions
are observed, are calculated from the Hessian matrix of the template
fit to the BDT response distributions.
Final results of the study are presented in
figure~\ref{comb_ilc}. As expected, the semi-leptonic event selection results
in the strongest limits, combining high event statistics (about 47\%
of decays) with background lower than for the hadronic channel.  
Including loose selection categories improves the expected limits by
20-30\% for the whole considered mass range.
When comparing to the decay independent limits, we can state that the
targeted analysis results in over order of magnitude increase in
sensitivity.
However, as the limit includes the branching ratio, we
can conclude that the di-tau channel is more sensitive only if the branching
ratio is of the order of 10\% or above.

\section{Search for $S \to b \bar{b}$}

\begin{figure}[h]
\centering
\includegraphics[width=0.97\twofigwidth]{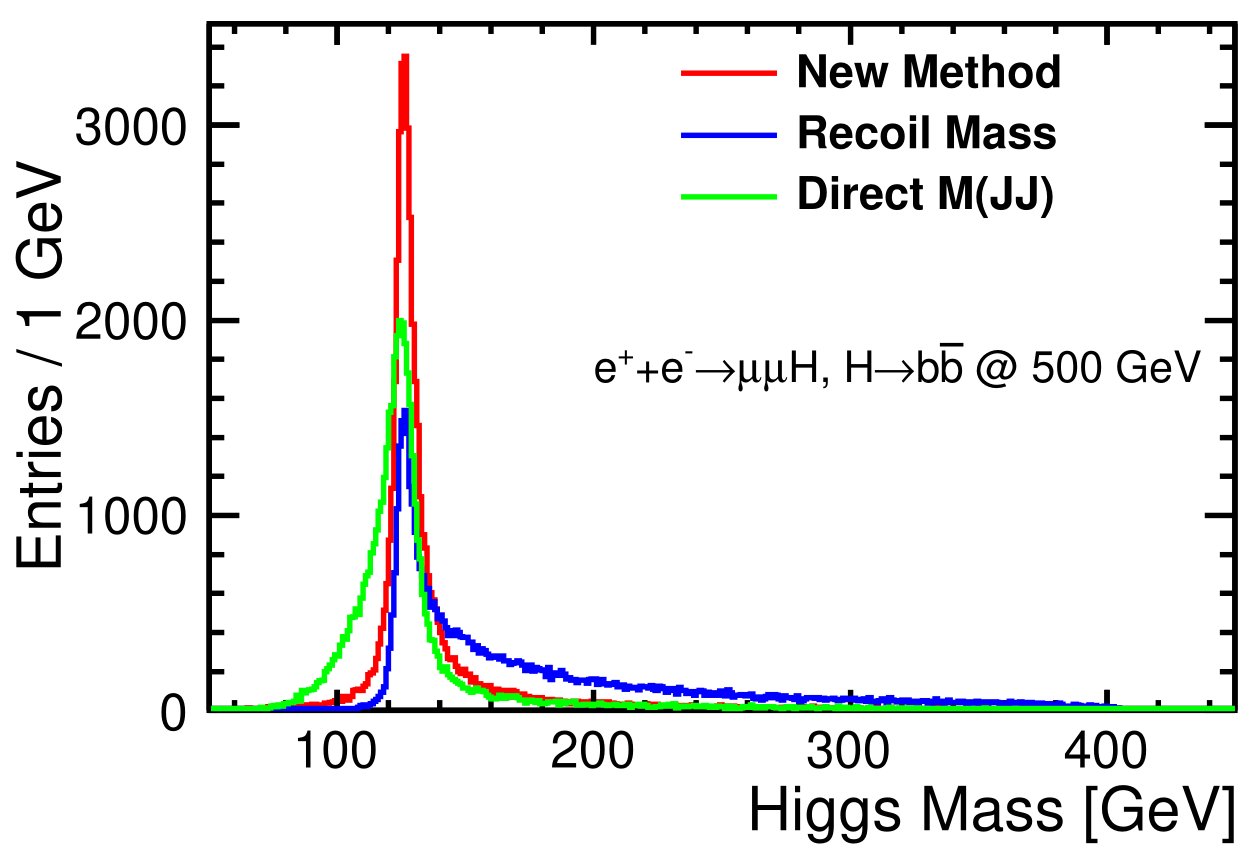}
\includegraphics[width=1.03\twofigwidth]{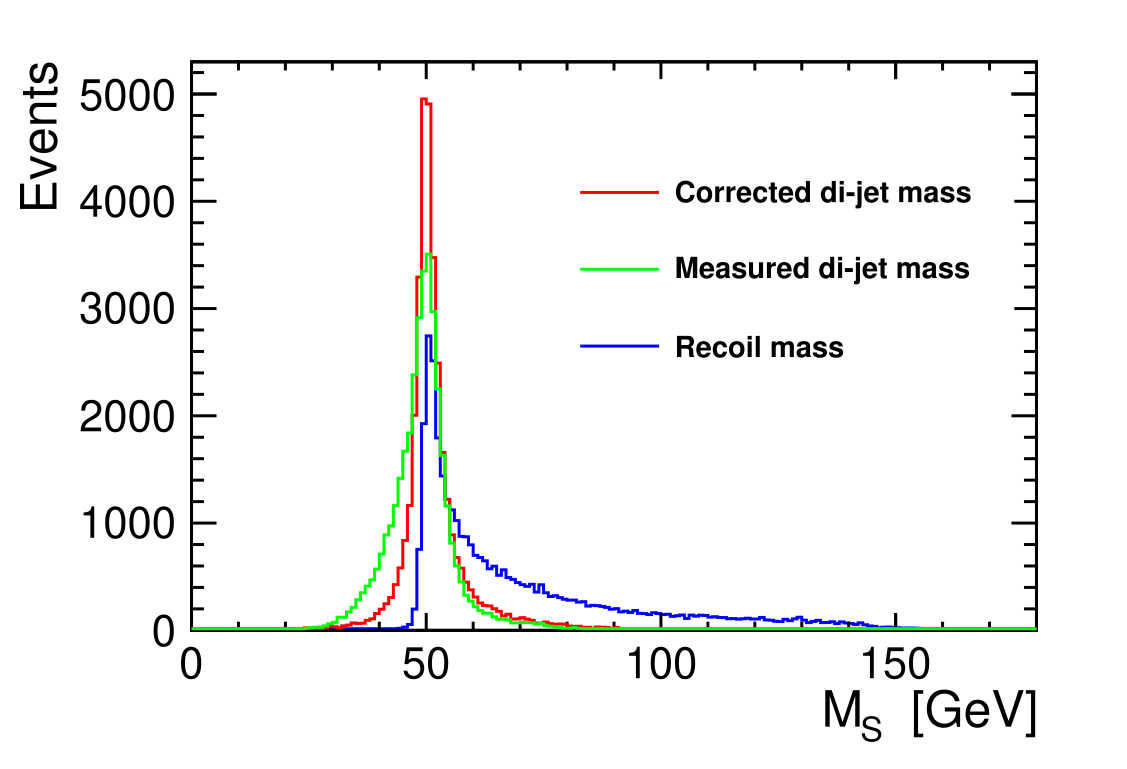}
\caption{Invariant mass of two jets reconstructed using raw jet
  energies, corrected jet energies and recoil method.
  Left: full simulation results for the 125\,GeV Higgs boson
  production at 500\,GeV ILC~\cite{bib:Tian}.
  Right: \delphes simulation results for the signal of 50\,GeV scalar
  production at 250\,GeV ILC.
}
\label{bb_rec}   
\end{figure}

If the structure of new light scalar couplings to SM particles is similar
to that of the SM Higgs boson then the decay to
$\textrm{b}\bar{\textrm{b}}$ is expected to dominate down to the
masses of the order of 10\,GeV.
As huge hadronic background is expected from pair production of W
bosons, we focus on leptonic Z boson decays in this search channel.
Again, invariant mass of the new scalar,
as reconstructed directly from the two b-quark jets, is poorly measured  due
to neutrinos escaping in semi-leptonic heavy meson decays.  
However, as leptons from Z decays can be very precisely measured, we
can  use conservation of transverse momentum to reconstruct jet
energies from leptonic final state and jet angles. 
This approach was first proposed for Higgs mass measurement at the
ILC~\cite{bib:Tian}, but we verified that it works very well also for
light scalar production.
This is shown in figure~\ref{bb_rec}, where reconstructed di-jet mass
distributions are compared before and after correction for 125\,GeV
Higgs boson and new scalar of 50\,GeV.
Mass reconstructed with the recoil method (based on the Z decay
measurement only) is also included for comparison.
One can clearly see that jet energy correction based on the transverse momentum
conservation significantly improves scalar mass measurement and allows
to obtain precision much better than with uncorrected jet energies or
the recoil method.
The study of light scalar production in $\textrm{b}\bar{\textrm{b}}$
decay channel is ongoing and the expected exclusion limits will be
presented at the 3rd ECFA workshop in Paris.

\section{Conclusions}

BSM scenarios involving light scalars, with masses accessible \epem
Higgs factories, are still not excluded by existing data.
Sizable production cross sections for new scalars can also coincide
with non-standard decay patterns, so different decay channels should
be considered.
Strong limits are already expected from decay independent searches,
based on recoil mass reconstruction.
These are expected to be improved further with new analysis methods
and the corresponding study is ongoing.
Compared to the decay independent search, more than an order of magnitude limit
improvement in search sensitivity is expected for light scalar decays
to tau pairs, if this is a dominant decay channel.
Studies of other decay channels are ongoing and the results are being
prepared for the ECFA study report.

\subsection*{Acknowledgments}

This work was carried out in the framework of the ILD concept group as
a contribution to the ECFA e$^+$e$^-$ Higgs/EW/top factory study.
This work was supported by the National Science Centre,
Poland, under the OPUS research project no. 2021/43/B/ST2/01778.


\printbibliography

\end{document}